\documentclass[journal=jacsat,manuscript=article]{achemso}

\usepackage[version=3]{mhchem} 
\usepackage{xcolor}
\usepackage{amsmath}
\usepackage{siunitx}
\usepackage{color,soul}
\usepackage{overpic}
\usepackage{float}
\usepackage{tablefootnote}
\usepackage{threeparttable}
\usepackage{ulem}

\author{Germaine Neza Hozana}
\affiliation[ICTP]{International Centre for Theoretical Physics (ICTP), Strada Costiera 11, 34151 Trieste, Italy}
\alsoaffiliation{Dipartimento di Fisica, Universit\'a degli Studi di Trieste, Via Alfonso Valerio 2, 34127, Trieste, Italy}
\author{Gonzalo Díaz Mirón}
\affiliation[ICTP]{International Centre for Theoretical Physics (ICTP), Strada Costiera 11, 34151 Trieste, Italy}
\author{Ali Hassanali}
\affiliation[ICTP]{International Centre for Theoretical Physics (ICTP), Strada Costiera 11, 34151 Trieste, Italy}
\email{ahassana@ictp.it}

\title[An \textsf{achemso} demo]
  {Data-Driven Discovery of the Origins of UV Absorption in Alpha-3C Protein}

\abbreviations{IR,NMR,UV}
\SectionNumbersOn

\begin{document}


\begin{abstract}

Over the last decade, there has been a growing body of experimental work showing that proteins devoid of aromatic and conjugated groups can absorb light in the near-UV beyond 300 nm and emit visible light. Understanding the origins of this phenomena offers the possibility of designing non-invasive spectroscopic probes for local interactions in biological systems. It was recently found that the synthetic protein $\alpha _3$C displays UV-vis absorption between 250-800 nm which was shown to arise from charge-transfer excitations between charged amino acids. In this work, we use data-driven discovery to revisit the origins of these features using molecular dynamics and excited-state simulations. Specifically, an unsupervised learning approach beginning with encoding protein environments with local atomic descriptors, is employed to automatically detect relevant structural motifs. We identify three main motifs corresponding to different hydrogen-bonding patterns that are subsequently used to perform QM/MM simulations including the entire protein and solvent bath with the density-functional tight-binding (DFTB) approach. Hydrogen-bonding structures involving arginine and carboxylate groups appear to be the most prone to near-UV absorption.  We show that these features are highly sensitive to the size of the QM region employed as well as to the inclusion of explicit solvation underscoring the limitations of previously used gas-phase cluster models.

\end{abstract}

\section{Introduction}

The electronic absorption spectra of many biological systems primarily appears within the ultraviolet region located between 180-320 nm\cite{lakowicz2006principles}. The common spectroscopic rationale is that chemical moieties characterized by aromatic or conjugated groups are responsible for these electronic transitions and subsequently, the emission of visible light through fluorescence. In the last decade, there has been a growing body of experimental work showing otherwise, namely that supramolecular assemblies constituting amino acids\cite{ARNON2021102695,Kumar2022}, peptides\cite{Shukla2004}, proteins\cite{chung2022label,pansieri2019ultraviolet,Pinotsi2016,pinotsi2013label}, and sugar polymers\cite{yu2019oligosaccharides} devoid of any classical conjugation, display the possibility of broad UV absorption ranging between 300-400 nm and fluorescence in the visible green-blue regime. More recently, there have also been reports of so-called cluster-triggered emission in a wide variety of supra-molecular assemblies\cite{wu2024multiple,xie2024cluster,liu2020mdm2,bresoli2024polymers}. Being able to harness these electronic spectroscopic features could open up a new class of non-invasive and non-toxic probes for biophysical processes in solution.

There have been numerous proposals to explain these observed phenomena including delocalized electrons along hydrogen bonds in peptide assemblies\cite{Shukla2004}, proton transfer along short hydrogen bonds\cite{stephens2021short,Pinotsi2016}, carbonyl group fluorescence\cite{Niyangoda2017carbonyl,Grisanti2020,miron2023carbonyl,morzan2022non} and charge-transfer excitations between polar amino acids\cite{prasad2017near}. How these different affects are manifested in different chemistries and whether they operate independently or simultaneously remains an open question. Given that these observed phenomena appear to challenge common chemical and spectroscopic intuitions, theory and computation has played an important role in trying to decipher the underlying mechanisms. 

In this work, we focus on the optical properties of a synthetic protein, $\alpha_3$C a three helix bundle protein where 50\% of the amino acids are charged. Recently, Prasad and co-workers showed using UV-vis absorption spectroscopy that $\alpha_3$C\cite{prasad2017near}, which is devoid of any aromatic or conjugated groups, shows a broad absorption between 250-800 nm. In addition, later experiments also found that $\alpha_3$C displays fluorescence in region 310-550 nm when excited at 295 nm \cite{Kumar2022}. Using excited-state quantum chemistry calculations of amino acids extracted from the protein, they found that the low-energy transitions between 250-800 nm could be attributed to charge-transfer transitions between the negatively charged carboxylate groups and backbone groups as well as positively charged lysines. These calculations were shown to quantitatively reproduce the the experimental spectra. But are vacuum conditions sufficient for reproducing the experimental trends?

Herein, we re-visit the optical properties of $\alpha_3$C combining techniques from unsupervised machine-learning with time-dependent tight binding density functional theory (TD-DFTB)\cite{niehaus2001tight,niehaus2009approximate}, to unravel the origins of the UV-vis absorption in the protein. Our group has recently shown that TD-DFTB does exceptionally well at reproducing structural and optical properties in amino-acid crystals\cite{Gonzalo2024}. The validity of the semi-empirical approach significantly expands the scope of applications to larger biological systems such as $\alpha_3$C. Using the smooth-overlap of atomic positions descriptors (SOAP)\cite{bartok2013representing,C6CP00415F} and density peak clustering\cite{DERRICO2021476}, we first identify structural motifs of the $\alpha_3$C protein that subsequently serve as input for QM/MM simulations from which TD-DFTB simulations are performed. 

Our procedure automatically identifies three structural motifs involving the amide-peptide backbone, lysine and finally arginine amino acid all of which form thermodynamically stable clusters involving hydrogen bonding interactions with other protein chemical groups as well as water molecules. These three clusters are then used to perform QM/MM simulations and subsequently excited state calculations. Interestingly, we find that hydrogen bonding interactions involving arginine and carboxylate groups seem to be the key interactions leading to charge-transfer excitations extending from 250-350 nm. The extent of this red-edge absorption is highly sensitive to the local chemistry and solvation included into the QM region. We show that TD-DFT calculations of amino-acid clusters in vacuum artificially lead to the presence of low-energy tails consistent with the experiment, albeit, for the wrong reasons. We propose that nuclear quantum effects (NQEs) might be the key ingredient for reproducing the experimental red tail absorption. We believe that our results provide an important general framework for understanding the optical properties of non-aromatic systems where design principles are critically needed to direct and interpret future experiments in the field.

\section{Computational Details}

\begin{figure}[H]
    \centering
    \includegraphics[scale=0.56]{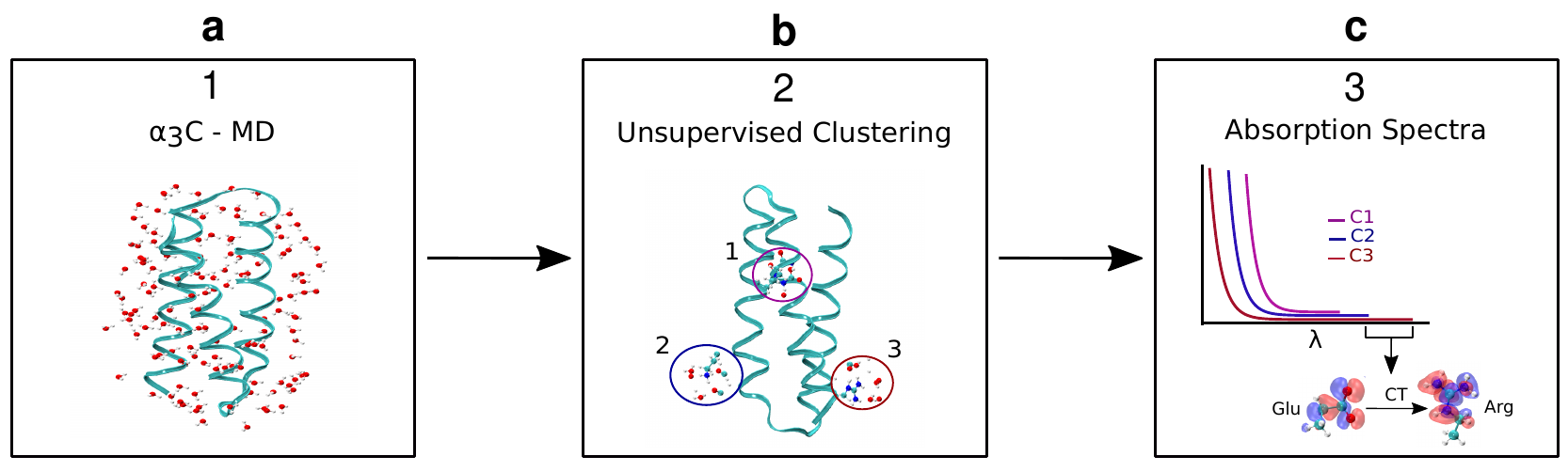}
    \caption{The three steps in our data-driven approach are summarized in the schematic. The first step (1) involves performing MD simulations of the $\alpha_3$C protein. Configurations obtained from the MD are then used to build local-atomic descriptors followed by dimensionality reduction and subsequently clustering (2). Finally, the identified clusters are then used perform calculations using TDDFTB to obtain absorption spectra.}
    \label{fig:protocol}
\end{figure}

In this section, we describe the computational protocols employed to study the optical properties of the $\alpha_3$C protein in aqueous solution. Our methodology involves several key steps, summarized in Figure \ref{fig:protocol}.

Initially, we conducted classical molecular dynamics (MD) simulations of $\alpha_3$C in water (1). The trajectory was then pipelined using machine learning techniques to automatically identify the relevant structural patterns (2). From these identified clusters, we performed Quantum Mechanics/Molecular Mechanics (QM/MM) simulations on the ground electronic state (3). Finally, we computed the absorption spectra for each relevant cluster, providing insights into the protein's optical properties. In the following, we will describe the specific details for each of these steps.

\subsection{Classical Molecular Dynamics} \label{md}

We began by constructing the system using the conformation of the $\alpha_3$C protein obtained through nuclear magnetic resonance (NMR) from the Protein Data Bank (PDB code 2LXY)\cite{tommos2013reversible}. The protein was then solvated in a rectangular box of water molecules with dimensions 6.7 $\times$ 5.6 $\times$ 6.0 nm$^3$, and two chloride ions were added to neutralize the system. For all classical simulations, we employed the GROMACS software suite\cite{lindahl2022gromacs}, using the CHARMM27 force field\cite{mackerell1998all} for the protein and the TIP3P model\cite{price2004modified} for water molecules as used in a previous study. 

Our simulation protocol began with 10000 steps of energy minimization to relax the system and remove any steric clashes or unfavorable contacts. This was followed by an initial 100 ps equilibration process involving a gradual rise in temperature from 0 K to 300 K using the NVT ensemble. Subsequently, a 200 ps equilibration at constant pressure and temperature (1 atm and 300 K) was performed using the NPT ensemble to equilibrate the density. The final production run consisted of a 1 µs simulation in the NPT ensemble, maintaining the same temperature and pressure. The V-rescale thermostat\cite{bussi2007canonical} was used to control the temperature, while the Parrinello-Rahman barostat\cite{parrinello1981polymorphic} was employed for pressure coupling.  
The protein maintained its three-helix bundle structure over the simulation period (see SI Figure \ref{fig:rmsd}).

\subsection{Unsupervised Learning Techniques}

Using configurations extracted from the MD simulation, we applied machine learning algorithms to identify the most relevant structural motifs that could serve as input for the optical calculations. This process involves three critical steps: firstly the use of a local atomic descriptor to characterize the local environments, secondly estimating the intrinsic dimension (ID) of the dataset and finally, a clustering technique to group similar environments together.

In this work, we used the Smooth Overlap of Atomic Positions (SOAP) descriptor\cite{bartok2013representing,C6CP00415F}, which has emerged as a powerful technique for encoding information about local environments in a wide variety of molecular systems, including organic molecules\cite{C6CP00415F}, biological systems\cite{maksimov2021conformational,lange2024comparative}, solid-state systems\cite{bartok2018machine,reinhardt2020predicting,helfrecht2019new}, and also in liquid water \cite{donkor2023machine,offei2022high,di2023zundeig,donkor2024beyond,capelli2022ephemeral,monserrat2020liquid}. The SOAP descriptor captures the spatial distribution of atoms around a central atom by considering both the type and position of neighboring atoms within a specified radius. Its key strength lies in its ability to create a smooth, continuous representation of the atomic environment that preserves important symmetries, allowing for detailed comparisons between different local structures. For a more detailed description of the construction of the SOAP descriptors, the reader is referred to previous literature\cite{bartok2013representing,C6CP00415F,himanen2020dscribe}. Here we give a brief summary of main ingredients.
Given an atomic environment $\chi$ around a central atom, one determines the local density as a sum of Gaussian functions with variance $\sigma^{2}$ centered on each of its neighbors including the central atom itself,

\begin{equation}
    \rho_{\chi} (r) = \sum_{i \in \chi} \exp{\left(-\dfrac{(x_i-r)^2}{2 \sigma^2}\right)}
    \label{eq1}
\end{equation}

This atomic neighbour density can be expanded in terms of radial basis functions and spherical harmonics $Y_{lm}$ yielding, 

\begin{equation}
    \rho_{\chi} (r) = \sum_{n=0}^{n_{max}} \sum_{l=0}^{l_{max}} \sum_{m=-l}^{l} c_{nlm}g_n(r) Y_{lm}(\theta, \phi)
    \label{eq2}
\end{equation}

Where the $c_{nlm}$ are the expansion coefficients. One can then construct a rotationally invariant power spectrum $\bf{p}$ with elements given by,

\begin{equation}
    p_{nn'l} = \pi \sqrt{\frac{8}{2l+1}} \sum_m (c_{nlm})^\dagger c_{n'lm} 
    \label{eq3}
\end{equation}

We put SOAP centers on all nitrogen atoms of the protein, and include in the environment, nitrogen and oxygen atoms regardless of whether they come from the protein or water molecule type. We employed a cutoff radius of 4 \AA, chosen based on the radial distribution functions between nitrogen and neighboring oxygen atoms (see SI Figure \ref{fig:rdf}). Our motivation for using this setup to build the chemical environments is dictated by the fact that the strong polar interactions in $\alpha_3$C involve N-H hydrogen bond donors either along the protein backbone or coming from the side-chains. Using the oxygens of the protein to build SOAP centers led to the identification of very similar statistically significant clusters. We have found that increasing the cutoff radius does not change the outcome of our results (see SI Figure \ref{fig:SI_clusters}). The width of the Gaussian function was set to 0.25 \AA, consistent with previous studies from our group\cite{donkor2023machine}. The maximum number of radial (nmax) and angular basis (lmax) functions was set to 8 and 6, respectively.

With the SOAP features in hand we employed the Two-NN estimator \cite{facco2017estimating}, a method used to estimate the ID based on information from the first and second nearest neighbors of data points. It was shown in the work of \citeauthor{facco2017estimating}, that the ratio of the second to the first nearest neighbor distances ($\mu = r_2 / r_1$) follows a specific distribution under the assumption of local uniform density namely,

\begin{equation}
P(\mu) = \frac{d}{\mu ^{d +1}}
\end{equation}

where $d$ is the ID. Assuming independence of sampled ratios $\mu_{i}$, the ID can be estimated through a maximum likelihood technique as:

\begin{equation}
d = \frac{N}{\sum_{i = 1}^{N}{\log(\mu_i)}}
\end{equation}
Where $N$ is the total number of samples in the dataset.

To construct the high-dimensional free energies, we employed the Point Adaptive \emph{k}-nearest neighbor estimator (PA\emph{k}). This technique uses the ID to construct a point-dependent density ($\rho_i$) which is determined by adding a linear term to the $k$-nearest neighbor estimator, where the density is $\rho_i = \frac{k_{i}}{r_{k_{i}}^{d}}$. The $k_i$'s are estimated for each data point as the larger neighborhood for which the density is approximately constant. The point-dependent free energy is then determined as $-Log(\rho_{i})$. The combination of these unsupervised techniques has been successfully used in our group for several other applications involving liquid water\cite{donkor2024beyond,offei2022high} and concentrated acids\cite{di2023zundeig}.

In our analysis, we randomly sampled 100000 data points from the SOAP dataset and performed DPA clustering using $Z=14$. Note that the Z-parameter in DPA determines the statistical confidence of the cluster. The sensitivity of our results to the choice of this parameter are discussed in the SI, see Figure \ref{fig:SI_clusters}. 
We use distance-based analysis of data-manifolds in Python (DADApy)\cite{dadapy}, a Python software library, for all the analysis.
For visualization purposes, we employed the Uniform Manifold Approximation and Projection (UMAP) method\cite{mcinnes2018umap}, to project the high-dimensional dataset into a 2D space. Additionally, we also performed the same analysis using the HDBSCAN clustering method\cite{ester1996density} for comparison and validation obtaining consistent results.

\subsection{QM/MM molecular dynamics simulations}

Following the identification of the different clusters in our system, 10 distinct configurations were randomly selected from each cluster to run the QM/MM ground state electronic simulations. Before delving into the details of these simulations, we will describe our protocol for partitioning the system into the QM and MM regions.

The output provided by our clustering technique identifies various environments around the hydrogen bonds involving the nitrogen atom and its bound proton. Therefore for the QM region, we included all atoms that are approximately within a radius of 4 \AA \ from the center of mass of these hydrogen bonds. In some cases, if we identified a charged amino acid at the boundary, it was also included into the QM region. Each QM region comprises of up a maximum of 100 atoms including both solvent and protein atoms. Hydrogen atoms were added as needed at the boundaries of the QM/MM to serve as link atoms\cite{morzan2018spectroscopy,field1990combined}. The remainder of the system was assigned to the MM region. To assess the sensitivity of our analysis to the role of including QM waters, we also repeated some of our simulations removing all the waters from the QM region (see Results and Discussion for details).

The QM/MM simulations were run using the electrostatic embedding scheme\cite{morzan2018spectroscopy}, in which the electrostatic potential of the MM region influences the QM region, providing a more accurate representation of the interactions between the QM and MM regions. This approach ensures that the electronic structure of the QM region is properly polarized by the surrounding MM environment. All the simulations were performed using the GENESIS software\cite{Jung2015,Kobayashi2017}, which handles the propagation of nuclei and all MM calculations. The classical force fields employed in this simulation are the same as those that were used previously for the MD. Each simulation was run for 5 ps with a timestep of 0.5 fs, using the canonical-sampling velocity-scaling thermostat\cite{huang2010cell} to maintain a constant temperature at 300 K. For the QM calculations, we employed the Density Functional Tight Binding (DFTB) theory as implemented in the DFTB+ package\cite{hourahine2020dftb+}, utilizing third-order corrections with the 3ob Slater-Koster parameters\cite{gaus2013parametrization}. For more methodological details, the reader is referred to previous literature on the topic\cite{senn2009qm,ryde2016qm}.

\subsection{Absorption Spectra Calculations}

For each of the previously run QM/MM ground state simulations, we selected 100 snapshots at 50 fs intervals and performed excited state calculations on these configurations. We used two different setups for these calculations: the QM/MM absorption spectra, where the MM region was included as point charges, and the QM vacuum absorption spectra, where all MM atoms were excluded from the calculation.

We utilized the DFTB+ software for these calculations, employing Tight-Binding Time-Dependent Density Functional Theory (TD-DFTB)\cite{niehaus2009approximate} with Long-Range corrections using the ob2 Slater-Koster parameters\cite{vuong2018parametrization}. For each conformation, a total of 30 excited states were calculated. When computing and displaying the spectra, each electronic transition was broadened using a Gaussian function with a width of 1 nm. 

To validate the quality of TD-DFTB, we selected some previously studied dimers\cite{prasad2017near} and performed calculations in vaccum using TD-DFT/CAM-B3LYP/6-311g(d,p) with the ORCA software\cite{neese2012orca}.

\subsection{Path-Integral Simulations}

In order to assess the role of nuclear quantum effects (NQEs) on the absorption spectra, we conducted some path-integral molecular dynamics simulations on the electronic ground-state for a selected cluster (see Results for more details).  We selected a QM system from our protein and ran a simulation with classical nuclei in a vacuum using DFTB+ software employing the same parameters as previously described. Path integral simulations for the same cluster in vacuum were simulated using DFTB+\cite{hourahine2020dftb+} and i-Pi\cite{ceriotti2014pi,kapil2019pi} software, with the PIGLET thermostat\cite{ceriotti2012efficient} and four beads. In a previous work we have shown that the effect on the absorption spectra comparing classical and quantum simulations is qualitatively captured using both 2 and 6 beads\cite{law2015role}. Both classical and quantum simulations were performed at a constant temperature of 300 K for 30 ps with a timestep of 0.5 fs. After completing the ground state simulations, we selected 100 equidistant frames and determined the absorption spectra using TD-DFTB\cite{niehaus2009approximate} using the same conditions as described earlier. All 4 beads of the path integral were used to compute the absorption spectra and therefore a total of 400 frames were used.


\section{Results and Discussion}

In this section, we present and discuss the results that emerge from our unsupervised clustering of the local environments in the protein and their subsequent role on determining the absorption spectra.

\subsection{Hydrogen-Bond Network Motifs}

\begin{figure}[H]
    \centering
    \includegraphics[scale=0.5]{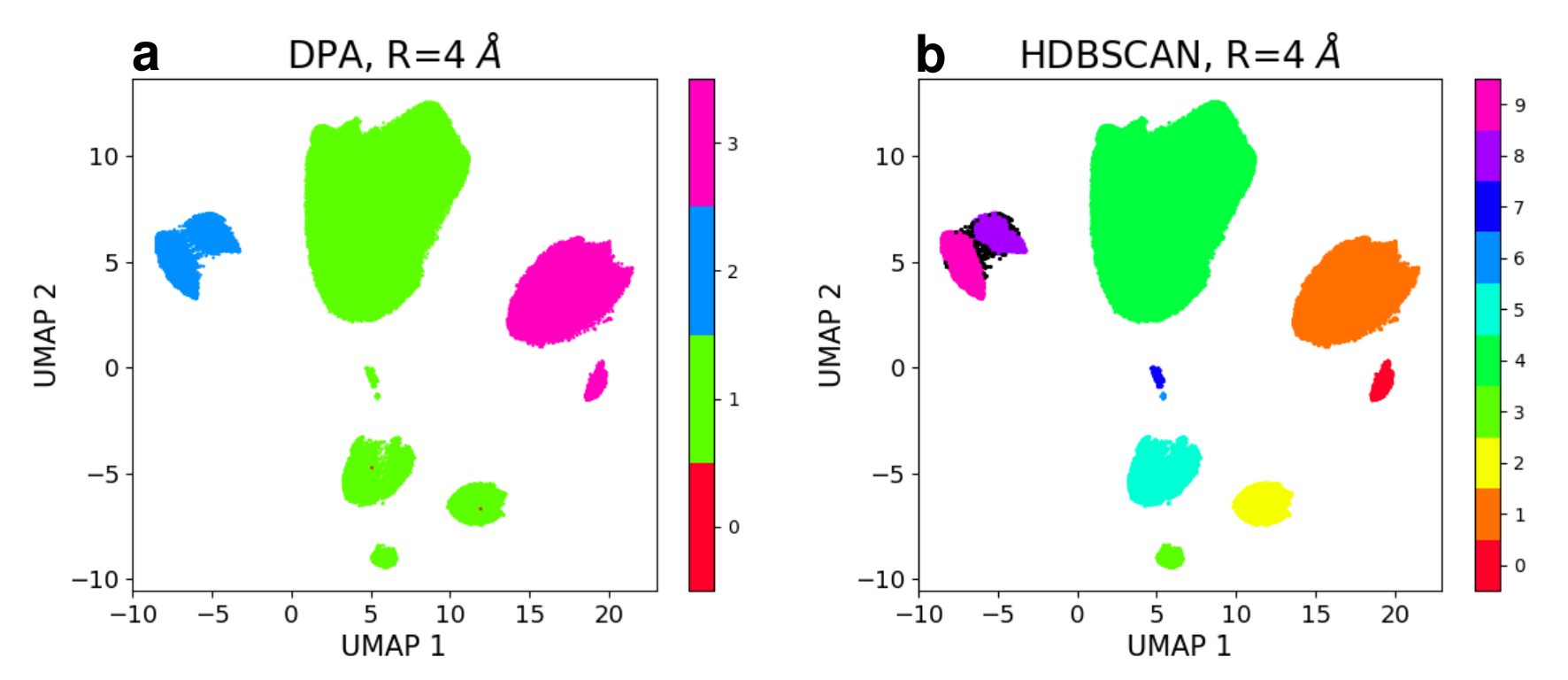}
    \caption{UMAP projections of the dominant clusters obtained using the DPA method (left) and HDBSCAN (right). The color codes correspond to the different clusters (4 in the case of DPA and 10 in the case of HDBSCAN).}
    \label{fig2}
\end{figure}

Our unsupervised clustering procedure in SOAP space yields three statistically dominant clusters. The left panel of Figure \ref{fig2} displays the UMAP projections colored based on the three dominant clusters for a SOAP cut-off radii of 4 \AA . The first cluster encompasses approximately 77\% of the regions, while the second and third clusters contain 16\% and 7\%, respectively. Less than 0.2\% of the total data-points are unclassified and are illustrated in the projection as black points. We have found that the existence of these three clusters is preserved for radial cutoffs that extend to larger values of 8 \AA \ as well as changing the Z parameter (see SI Figure \ref{fig:SI_clusters}). Overall, the three clusters consistently occupy over 99\% of the population. The right panel of Figure \ref{fig2} shows the outcome of the clustering using another clustering method namely, HDBSCAN which yields a total of 9 clusters. However, three of the clusters essentially occupy 83\% of the population similar to DPA.


\begin{figure}[H]
    \centering
    \includegraphics[scale=0.65]{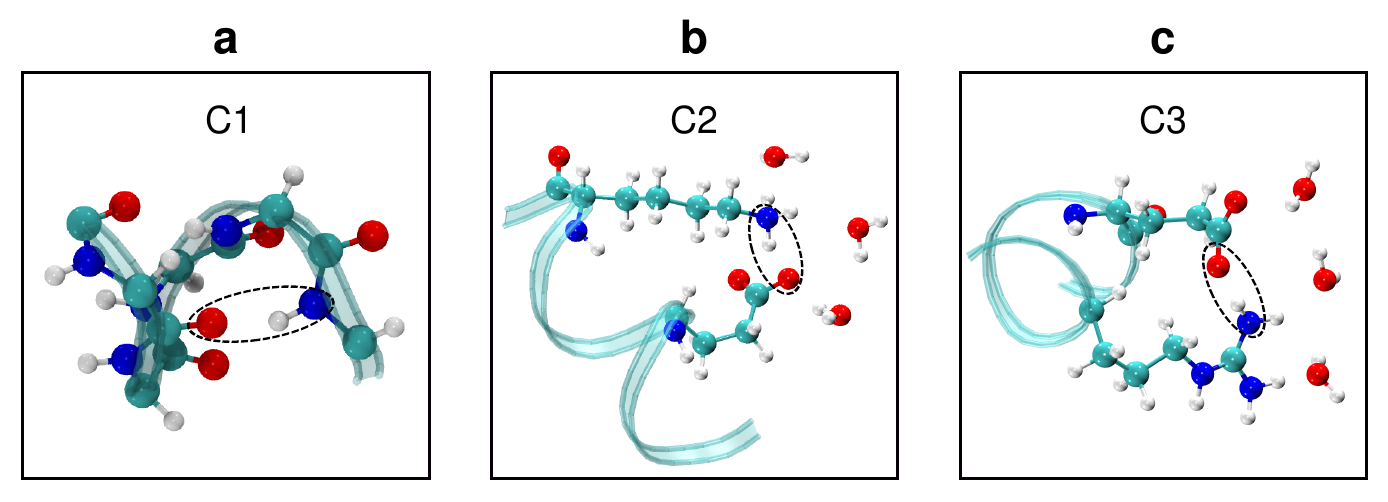}
    \caption{Schematic snapshots of the three clusters obtained from our analysis. Panel a) shows the backbone of the protein which primarily involves hydrogen-bonding of the amide-bonds along the helix (cluster-1, C1) . Panel b) shows lysine and glutamic acid hydrogen bonded to each other including some water molecules (cluster-2, C2). Panel c) shows arginine and glutamic acid side chains hydrogen bonded to each other also including some solvation water (cluster-3, C3).}
    \label{fig3}
\end{figure}

Having identified three statistically dominant clusters in the protein, we turned next to interpreting their chemical origin. As already eluded to earlier, since $\alpha_3$C consists of three helices, one might expect to see strong hydrogen bonds along the helix. Indeed, the largest cluster referred to in the rest of the manuscript as C1, consists of regions centered on nitrogen atoms of the protein backbone. These nitrogen atoms form intra-chain hydrogen bonds with oxygen atoms on the protein backbone (see leftmost panel of Figure \ref{fig3}) and occasionally also with water oxygens. 

Cluster-2 (C2) the second largest cluster, is primarily composed of regions centered on nitrogen atoms on the side chains of lysine residues (middle panel of Figure \ref{fig3}). A small but non-insignificant proportion of this cluster consist of nitrogen atoms at the N-terminus of the protein, particularly on glycine, the first residue. These nitrogen atoms are highly exposed to the  solvent and form hydrogen bonds with the side chains of glutamic acid residues and water molecules. Finally cluster-3 (C3) is dominated by the side-chains of arginine (rightmost panel of Figure \ref{fig3}). These nitrogen atoms are also exposed to the solvent forming hydrogen bonds with the side chains of glutamic acid and water. The chemical origin of the three dominant clusters obtained with HDBSCAN are fully consistent with these preceding findings.

\subsection{Absorption Spectra}

As eluded to earlier in the Introduction, Prasad and co-workers recently examined using UV absorption spectroscopy the optical properties of the $\alpha_3$C protein in solution where a broad UV absorption between 250-800 nm was identified\cite{prasad2017near}. In order to interpret the physical origins of this long-tail absorption, they conducted time-dependent density functional theory (TD-DFT) calculations in vacuum using clusters extracted from the protein. Focusing specifically on hydrogen-bonding interactions involving the charged amino (NH$_{3}$$^{+}$) and carboxylate (COO-) groups coming from the lysine and glutamic-acid side chains, they found charge-transfer excitations between these moieties that ranged between approximately 300-800 nm. The magnitude and intensity of these transitions was found to be rather sensitive to specific geometrical and environmental effects. For example, they found that pairs of charged amino acids that were positioned further away from each other ($\sim$5-6\AA) were more likely to lead to lower energy excitations. In addition by including water molecules into the cluster calculations, they found that some transitions could be blue shifted by $\sim$100-150nm while in other cases, adding other side-chains could enhance the intensity of the low energy transitions above 300 nm. 

\begin{figure}[H]
    \centering
    \includegraphics[scale=0.6]{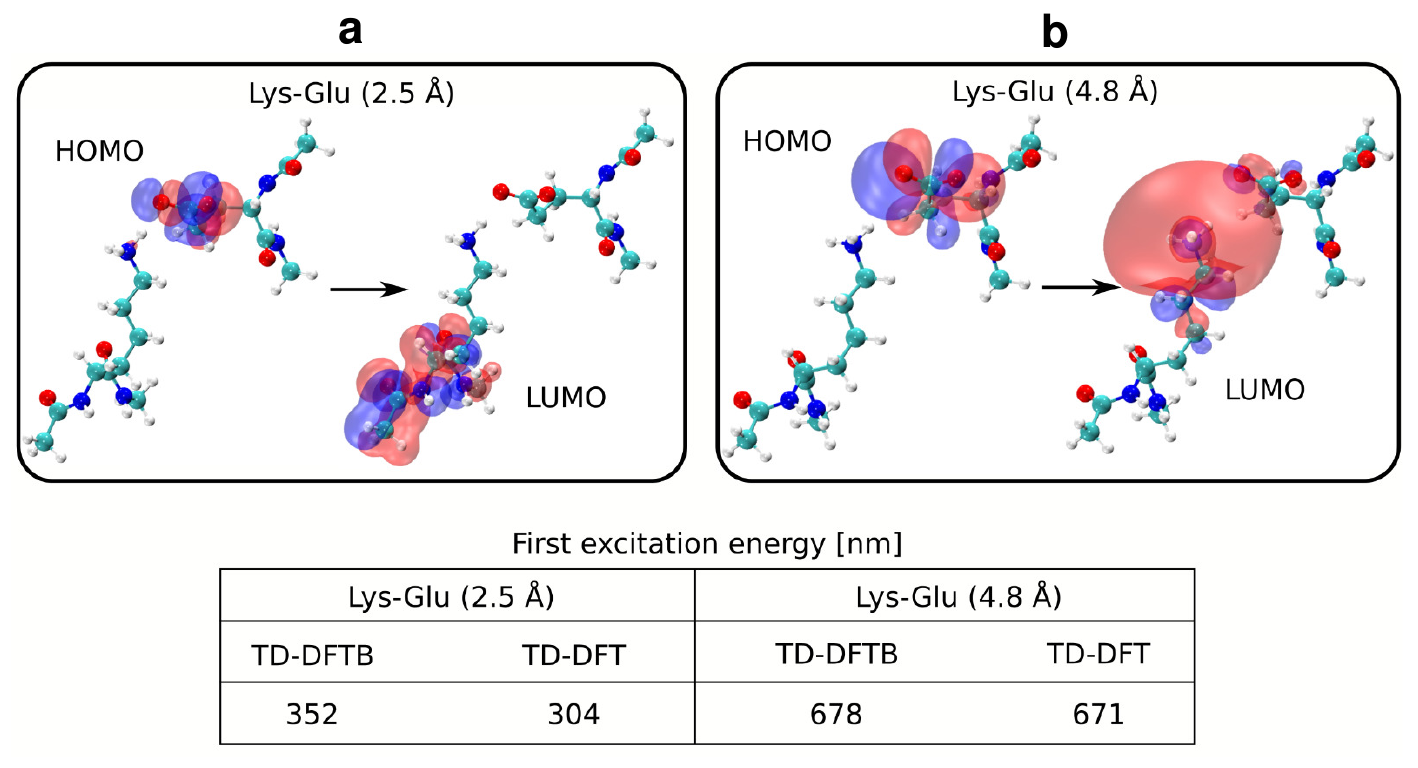}
    \caption{Comparison of first excitation energies between TD-DFTB and TD-DFT for two characteristic Lys-Glu dimers at distances of 2.5 \AA \ (a) and 4.5 \AA (b). The upper panel illustrates the molecular orbitals involved in the charge transfer transition. The lower panel provides a numerical comparison of the first excitation energies (in nm) calculated using TD-DFTB and CAM-B3LYP for the two dimers.}
    \label{fig:validation}
\end{figure}

The preceding results were conducted using TD-DFT with the range-separated functional CAM-B3LYP\cite{yanai2004new}. While this in principle provides a more accurate treatment of the electronic structure, it is computationally expensive and therefore limited to small system sizes. Our approach of using TD-DFTB overcomes these challenges but due to its semi-empirical nature needs to be first bench-marked to ensure that it reproduces the electronic structure obtained from TD-DFT. Figure \ref{fig:validation} compares the excitation energies and their contribution in terms of the molecular orbitals involved in the transitions for two different geometries that are simulated in vacuum using TD-DFT/CAM-B3LYP/6-311G(d,p) and TD-DFTB with Long Range corrections. These geometries involve the side-chains of lysine and glutamic acid forming a hydrogen bond at two different distances. For both methods, we observe an excitation energy ranging between 300-700 nm involving a charge-transfer excitation. These findings on the the gas-phase cluster model systems are thus fully consistent with the previous reported work by Prasad and co-workers\cite{prasad2017near}.

Having validated the TD-DFTB which reproduces previous findings, we are now in a position to move to applying it to the full protein. With the three clusters in hand determined in a completely agnostic manner we examined how these different motifs modulate the absorption spectra. We thus extracted the spectra conducting QM/MM simulations with DFTB on the electronic ground state and then determining the excited states of the system where short-range interactions with nearby protein and water moieties were included in the QM region, while the long-range interactions were treated by explicitly including the protein and water groups with classical electrostatics.

\begin{figure}[H]
    \centering
    \includegraphics[scale=0.4]{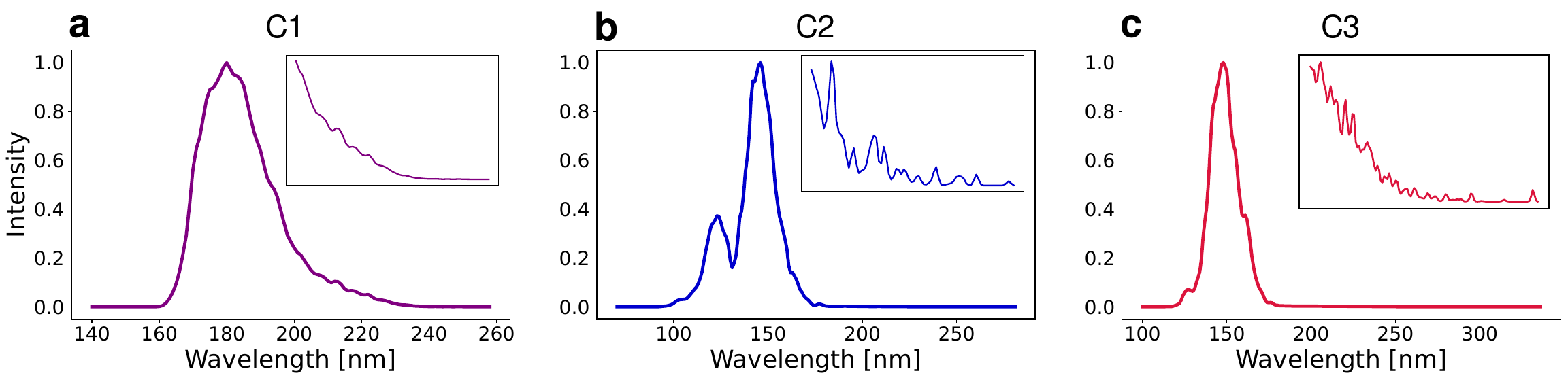}
    \caption{Absorption spectra obtained from the three clusters  with inset showing a zoom-in of the transitions above 200 nm. Out of the three clusters, C3 is the only one that displays transitions between 300-350nm.}
    \label{fig4}
\end{figure}

Figure \ref{fig4} compares the absorption spectra obtained for the three clusters averaging over a total of 1000 independent snapshots. At first glance, Figure \ref{fig4} appears to show that the spectra from all three clusters are dominated by a high intensity transition below 200 nm. In particular, C1 is peaked at 180 nm, C2 is characterized by a bi-modal structure involving a peak at 123 and 146 nm and finally, C3 has a single peak at 148 nm. However, upon closer examination C1 and C2 feature a tail in the absorption extending up to 280 nm as seen in the inset plots. It thus appears that the peptide backbone interactions and those involving lysine forming hydrogen bonds with the carboxylate when embedded in a realistic environment, no longer feature low-energy excitations below 300 nm. 

In contrast, C3 appears to be the only case where one observes a tail beyond 300 nm with a few some transitions near 340 nm. These transitions albeit weak, indicate that the hydrogen bonds involving arginine appear to be the only ones that introduce electronic states that appear in the mid UVA region. \emph{Thus the realistic inclusion of both the water and protein environment through a QM/MM setup does not lead to a long-tail of UV-absorption in sharp contrast to what is observed for the gas-phase clusters.}

\begin{figure}[H]
    \centering
    \includegraphics[scale=0.6]{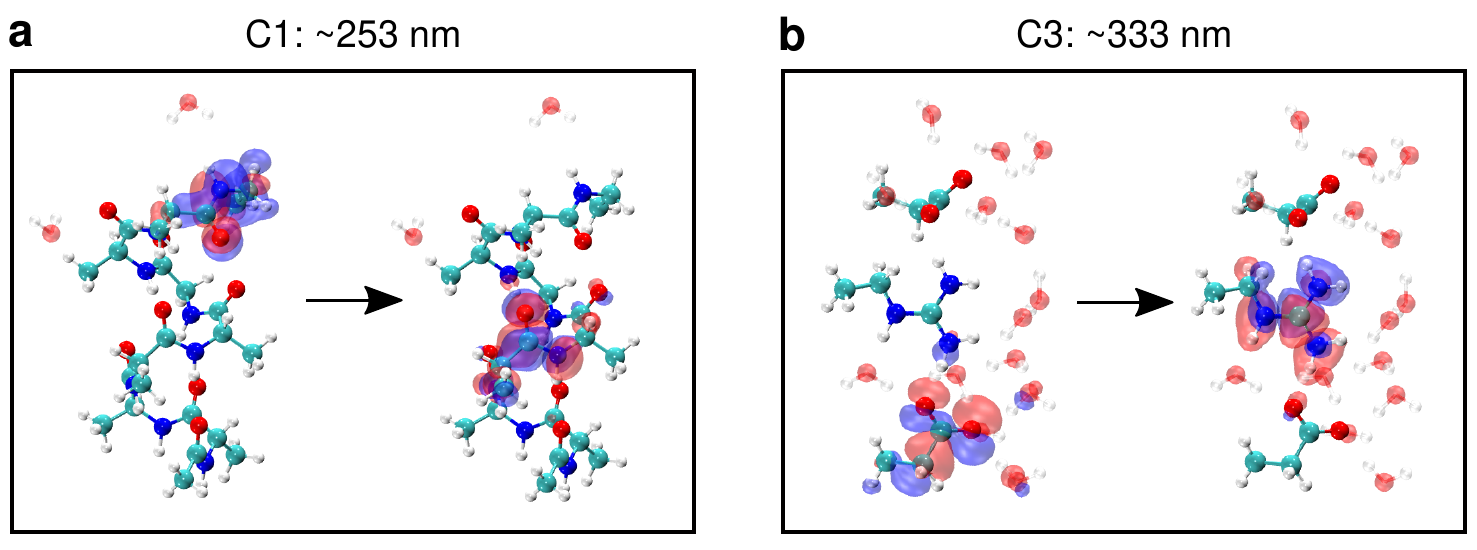}
    \caption{Low energy transitions in the first and the third cluster. The left panel a) shows a charge transfer from one amide group to another in C1, and the right panel b) shows a charge transfer from a glutamic acid carboxylate group to a guanidinium group of arginine in C3.}
    \label{fig5}
\end{figure}

In order to understand the electronic origins of the low energy tails of the spectrum for each cluster we examined the molecular orbitals (MO) associated with the lowest energy excitations as shown in the two panels of Figure \ref{fig5}. The electronic transitions in C1 correspond to a charge transfer $n$ to $\pi^*$ excitation from an amide-backbone group of one amino acid to another. In C2 instead, the low energy transitions correspond to a localized $n$ to $\pi^*$ excitation on the carbonyl group of the glutamic acid side-chain. Finally, the transitions higher than 300 nm found in the tail of C3 involves a HOMO-LUMO $n$ to $\pi ^*$ transition from the carboxylate group of glutamic acid to the guanidinium side-chain of arginine. 

\begin{figure}[H]
    \centering
    \includegraphics[scale=0.58]{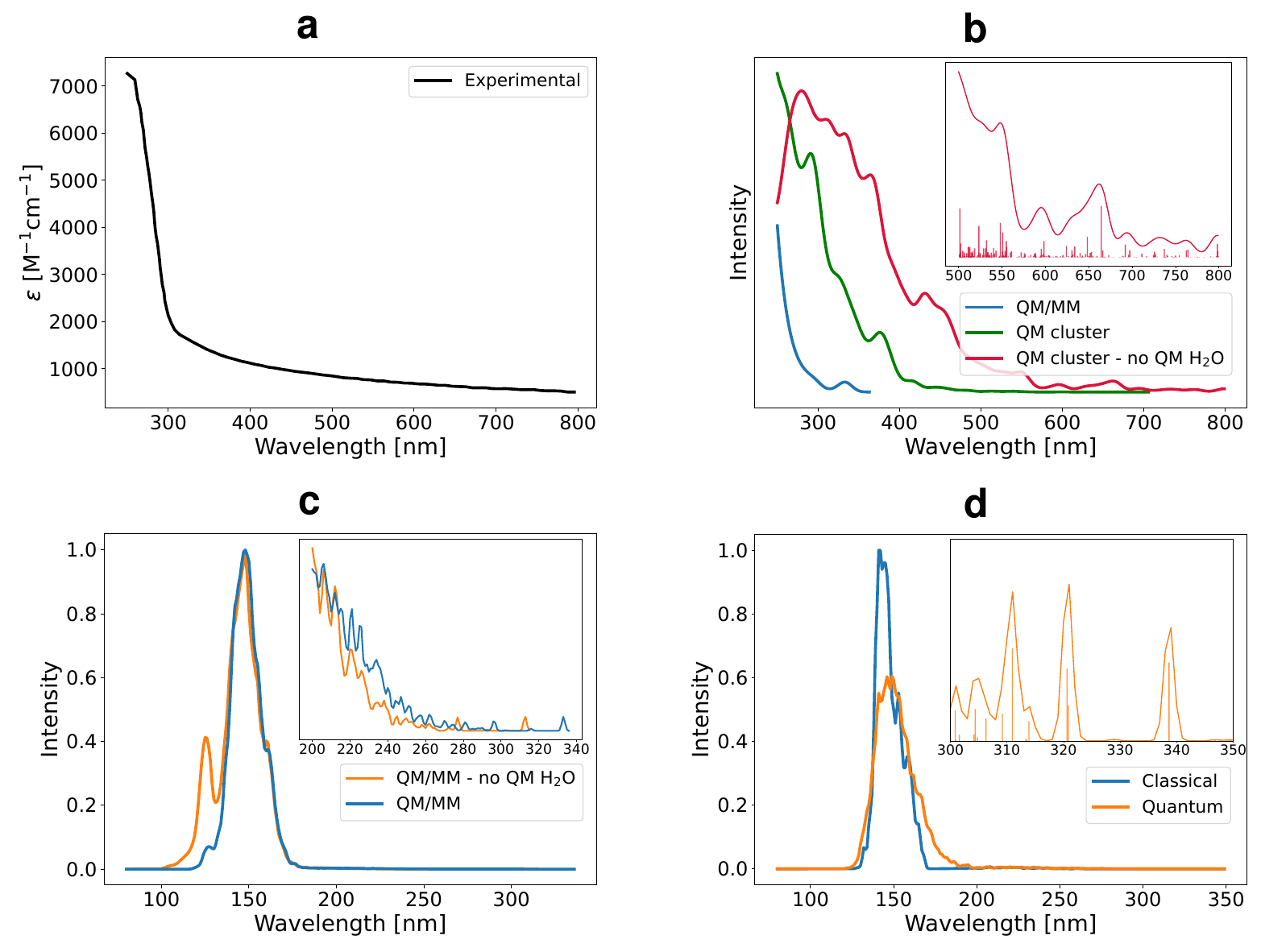}
    \caption{Panel (a) show the experimental absorption spectrum of $\alpha_3$C ranging from 250 nm, with the tail extending to 800 nm, adapted from \citeauthor{prasad2017near}. Panel (b) shows cluster-3 simulated absorption spectra, smoothened with a Gaussian of 10 nm width, calculated in three ways: 1) using our normal QM/MM setup described in the methods (solid blue curve), 2) using a QM cluster in vacuum (solid green curve) and 3) on QM cluster devoid quantum water molecules (solid red curve). The inset shows absorption spectrum for case 3) and corresponding excitations. Panel c) compares the absorption spectra from the full QM/MM protocol with and without the inclusion of water molecules in the QM part. Finally, panel d) shows the absorption spectra comparing classical and quantum simulations that show the role of nuclear quantum effects.}
    \label{fig6}
\end{figure}

The preceding results show that absorption spectra for all three clusters do not present any excitations above 350 nm. This observation is apparently in sharp contrast to that of the experiments which displays a long tail of absorption extending from 400-800 nm (see Figure \ref{fig6}(a). Over the years, there have been several studies showing the importance of solvation, specifically, the inclusion of explicit water molecules\cite{parac2010qm,nakayama2013solvent,improta2016quantum,sappati2016nuclear,law2015role} in tuning the optical absorption of organic molecules. Since our studies here involve a full QM/MM framework, we are in a unique position to examine the relative role of QM vs MM waters/protein in tuning the optical absorption.

Panel (b) of Figure \ref{fig6} shows the absorption spectra obtained from our normal QM/MM setup described earlier, to two other situations: i) the first involves carving out the QM region and converting it into a cluster in vacuum with the appropriate capping using hydrogen atoms which can in principle also contain QM water molecules (solid green curve) and ii) same as i) but now without including any waters into the QM cluster (solid red curve). Moving from the full QM/MM to scenario ii) results in a significant red-shift of 200 nm and furthermore introduces a long-tail in the absorption between 500-800 nm. Although this feature puts the theoretical predictions into closer agreement with the experiments, this feature appears to be an artifact of an unrealistic inclusion of the environment. Upon inclusion of water molecules into the QM region of the cluster, there is a blue-shift of approximately 100 nm moving the first excitations higher up in energy.
On the other hand, unlike the QM clusters scenario, removal of water molecules in QM region, in QM/MM settings, led to a slight blue-shift in the absorption spectrum of about 20 nm as shown in Figure \ref{fig6}(c).

All in all, our findings show an important and critical challenge of comparing absorption spectra from experiments with theory. On the one hand, it is possible to construct a model that almost quantitatively reproduces the experimental trends, in this case a less realistic QM cluster in vacuum. However, this consistency appears to be right for the wrong reasons - building systems that in principle account for environmental effects such as the protein and water, introduces bigger discrepancies between the theoretical predictions and experiments. It should however be stressed that our QM/MM approach appears to correctly capture an important region of the spectrum between 300-350nm that does not arise from aromatic groups.

In the last decade there has been a growing body of theoretical work showing the importance of including nuclear quantum effects (NQEs) into molecular simulations\cite{markland2018nuclear}. Specifically, zero-point energy (ZPE) fluctuations have been shown to affect structural\cite{ceriotti2013nuclear,ceriotti2016nuclear}, dynamical\cite{ceriotti2016nuclear,santoro2021quantum} and electronic\cite{sappati2016nuclear,law2015role,law2018importance,berrens2024nuclear,ambrosio2016structural,mondal2023effect,turi20242} properties of hydrogen-bonded systems. Due to the larger structural distortions induced by enhanced fluctuations of vibrational coordinates, absorption spectra can display substantial red-shifts in energy introducing lower-energy excitations that are completely absent in classical simulations.

In order to assess the sensitivity of our results to the role of NQEs we conducted path-integral simulations coupled with a colored noise thermostat (PIGLET) focusing on trying to understand how NQEs would shift energies in cluster-3 which displayed the lowest energy transition. Due to the computational cost, we could only conduct this analysis on the QM cluster in vacuum with a few solvating water molecules. Figure \ref{fig6}(d) compares the absorption spectra obtained from the classical runs to PIGLET. In the latter, the lowest energy excitation only reaches 260 nm while in the case of the PIGLET runs this extends to 340nm. Thus NQEs lead to a red-shift in the spectra of approximately 80 nm. Given these findings, we propose that the inclusion of NQEs could induce a further red-shift in our spectra reported earlier (Figure \ref{fig6}(d) which may bring the excitation down to the range of 400-500 nm.

\section{Conclusions and Perspectives}

There is currently a very active area of research from both experimental and theoretical fronts looking into the spectroscopic origins of biological systems that display a tendency to absorb UV radiation above 300 nm. A better fundamental understanding of this phenomena holds the promise to allow for designing non-invasive probes for biophysical processes. In this work, we have focused our efforts on studying the optical properties of $\alpha_3$C a synthetic protein which was recently shown to display near UV-visible absorption creating a long-tail of excitations between 300-800 nm. These previous studies conducted computational studies on gas-phase clusters which were shown to reproduce the experimental observations. 

Here instead, we employ a data-driven approach where state-of-the-art unsupervised learning approaches are used to automatically discover statistically important structural motifs in the protein. These are then used to conduct QM/MM simulations and optical absorption calculations, where the environment involving both protein and water molecules are included. This is achieved through the use of a tight-binding approach which we have recently demonstrated to give an excellent compromise between computational efficiency and accuracy for studying electronic properties of hydrogen-bond networks.

Our results leads to a somewhat surprising conclusion namely that a realistic inclusion of both the protein and water environment completely eliminates the long-tail of optical absorption between 400-800 nm. Instead, our approach correctly captures features between 300-350 nm which arise from charge-transfer excitations between arginine and carboxylate groups. We propose by conducting proof-of-concept path-integral simulations that one can expect an additional red-shift of up to 100 nm when quantum effects are included which would lead to a red-tail absorption extending to approximately 450nm. However, it remains an open question as to the source of discrepancy between the experimental observations and theoretical predictions. At this juncture, our theoretical predictions suggest that new experiments probing this red-tail in $\alpha_3$C are needed to eliminate the possibility of scattering artefacts and in addition by performing site-mutagenesis studies where the red-tail can be tuned.

\begin{acknowledgement}

G.D.M and A.H. acknowledge the funding received by the European Research Council (ERC) under the European Union’s Horizon 2020 research and innovation programme (grant number 101043272 - HyBOP). Views and opinions expressed are however those of the author(s) only and do not necessarily reflect those of the European Union or the European Research Council Executive Agency. Neither the European Union nor the granting authority can be held responsible for them. GDM and AH also acknowledge CINECA supercomputing (project NAFAA-HP10B4ZBB2) and MareNostrum5 (project EHPC-EXT-2023E01-029) for the resources allocation.

\end{acknowledgement}

\begin{suppinfo}
\setcounter{figure}{0}
\renewcommand{\thefigure}{S\arabic{figure}}

\subsection*{Structural Stability of $\alpha_3$C over 1 $\mu$s MD Trajectory}
\begin{figure}[H]
    \centering
    \includegraphics[scale=0.6]{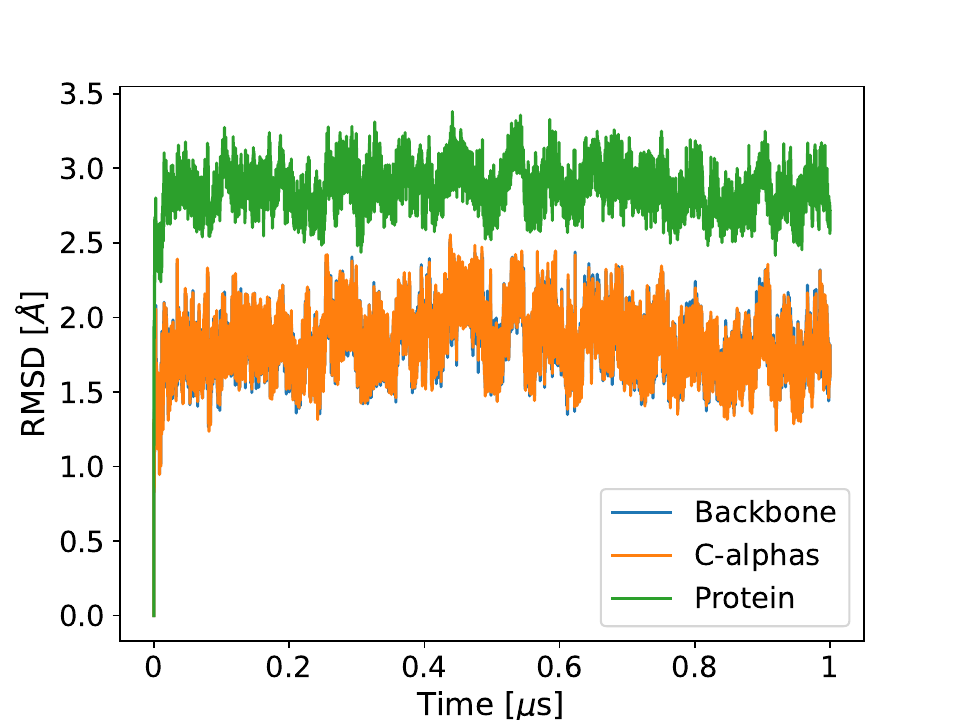}
    \caption{Root mean square displacement of $\alpha_3$C calculated over a one microsecond trajectory.}
    \label{fig:rmsd}
\end{figure}

\subsection*{SOAP Cutoff Radius}
\begin{figure}[H]
    \centering
    \includegraphics[scale=0.5]{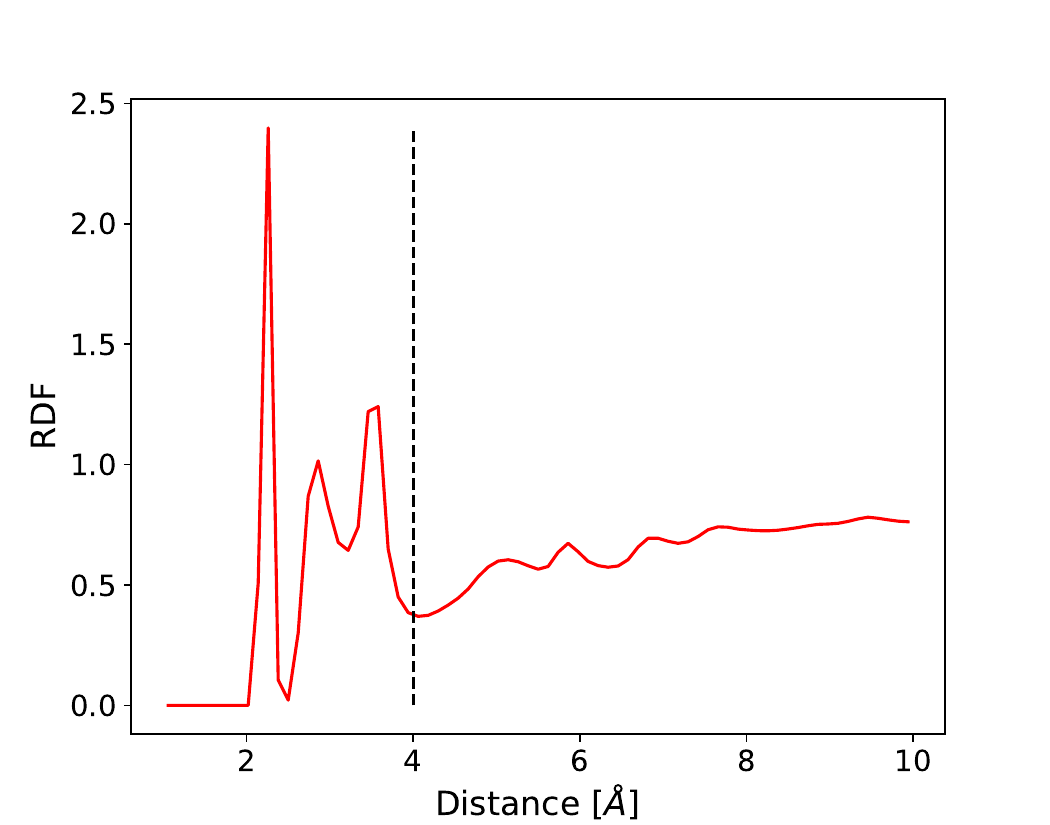}
    \caption{Radial distribution function between all nitrogen atoms in $\alpha_3$C protein and all oxygen atoms in the system. We chose $R_{cut}$=4 \AA \, the value below which all the sharp peaks were found.}
    \label{fig:rdf}
\end{figure}

\subsection*{ADP Clusters for Various $Z$ and $R_{cut}$ Parameters}
\begin{figure}[H]
    \centering
    \includegraphics[scale=0.56]{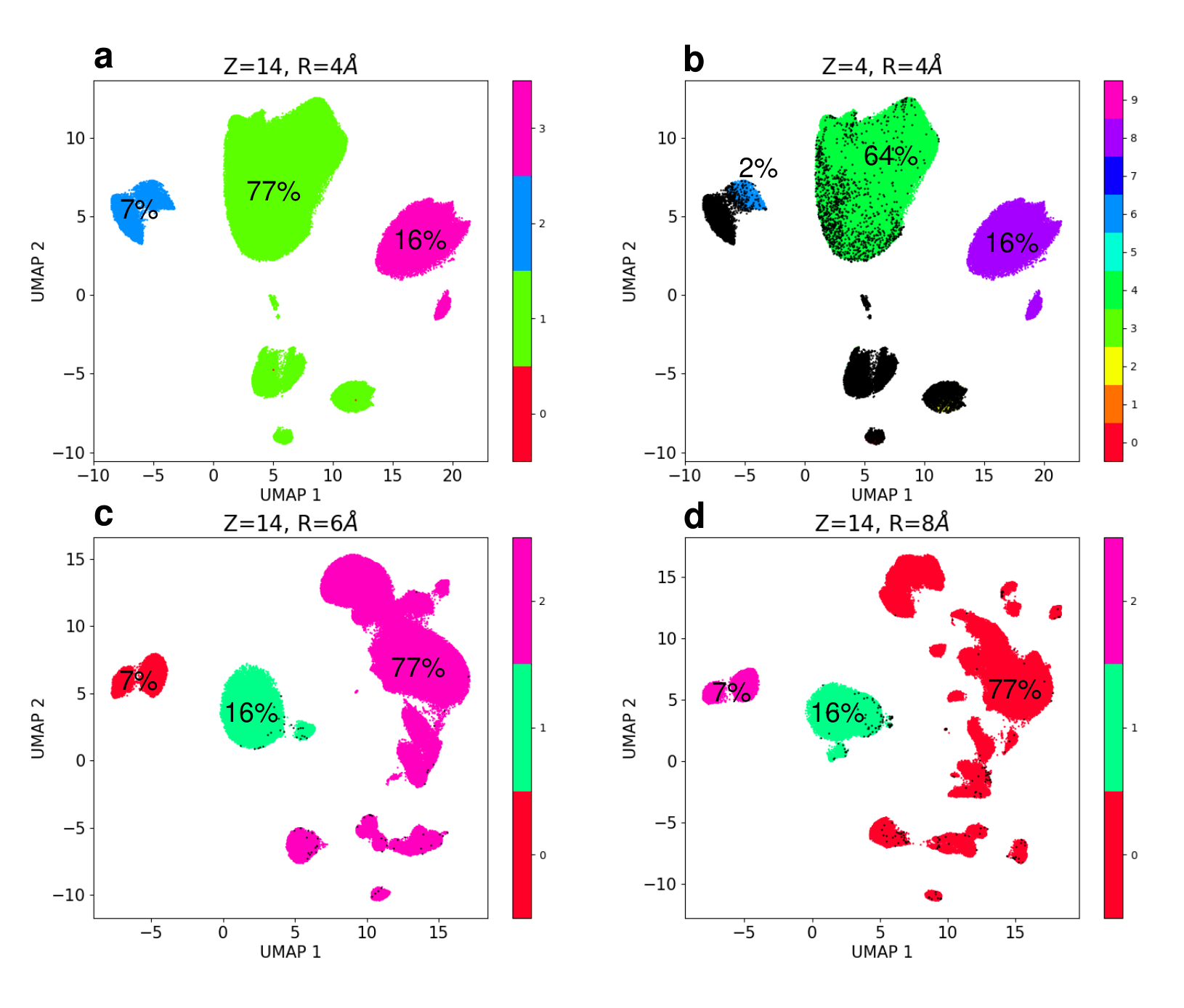}
    \caption{ADP clusters for various $Z$ and $R_{cut}$ values. Panel a: $Z=14$ and $R_{cut}=4$\AA, panel b: $Z=4$ and $R_{cut}=4$\AA, panel c: $Z=14$ and $R_{cut}=6$\AA, and panel d: $Z=14$ and $R_{cut}=8$\AA.}
    \label{fig:SI_clusters}
\end{figure}

\end{suppinfo}


\section*{Data Availability}
Data and the code used to produce figures presented in this article are available at \url{https://zenodo.org/records/13908181}.


\bibliography{references}

\end{document}